# Supercritical Snapping and Controlled Launching via Dual Latch Gels

Xiaona M. Xu[1], Nolan A. Miller[1], Gregory M. Grason[1*], Alfred J. Crosby[1*]

**Affiliations:**

[1]Department of Polymer Science and Engineering, University of Massachusetts Amherst, 120 Governors Dr, Amherst, MA, 01003-9263, USA

*Corresponding authors. G.M.G (grason@umass.edu) and A.J.C. (acrosby@umass.edu)

**Abstract**

Natural organisms have evolved integrated Latch-Mediated Spring Actuation systems (LaMSA) that consist of multiple latches and springs to enhance power output and adapt to diverse environmental conditions. Similar designs are appealing yet largely unexplored in engineered materials due to the complexity of integrating multiple components into a single material platform. Here, we report a dual-latched magneto-elastic shell device capable of selectively activating the latches to regulate snapping pathways and energy output based on specific actuation requirements. Differential deswelling across the thickness acts as the "motor" to load the elastic energy into the shell, which is then released via the snap-through instability once the loading reaches the critical threshold, constituting an intrinsic mechanical latch. Activation of the external magnetic latch delays snapping onset beyond the threshold of the intrinsic latch, leading to a power-amplified *supercritical* snap-through instability as well as a bifurcation instability. The combined function of both latches allows for flexible control over energy storage and release. Additionally, this integrated LaMSA system possesses an untethered anchoring mechanism, enabling the device to launch in arbitrary directions from the substrate, driven by the energy released during snapping. We envision that the design principles of dual-latched LaMSA systems will create opportunities for power-dense actuation in engineered materials and robotic devices.

**Main text**

Latch-mediated Spring Actuation (LaMSA) systems, which generate power-dense movements using a small amount of elastic energy[1, 2], exist in a wide range of organisms, from fungi to mantis shrimp[3-5]. In a LaMSA system, the latch is an opposing force that resists movements while the motor loads energy into the spring. The sudden removal of the latch then enables a rapid release of the stored energy, leading to amplified power output. The bistable leaf of *Dionaea muscipula* (Venus flytrap), which undergoes a rapid transition between two stable configurations via a snap-through instability to catch prey, exemplifies a LaMSA system in which the instability threshold serves as the intrinsic latch and turgor pressure functions as the motor[6-11]. This mechanism has increasingly guided the design of synthetic materials that mimic the power-dense snapping of Venus flytrap[12-17]. However, such intrinsic mechanical latch-based instabilities have limited control, resulting in snapping immediately upon reaching the critical internal stress threshold. Therefore, these systems lack independent control over snapping onset and energy release, constraining their adaptability across different applications. Other natural organisms have evolved integrated LaMSA systems that consist of multiple latches to enhance power output and control. Mantis shrimp, for example, have two latches to provide enhanced performance[2]. A small muscle-based latch holds the high impact-capable system in place until a virtual geometric latch is engaged and subsequently released to provide power-dense spring actuation[4, 18]. This integration allows mantis shrimp to adjust their latches depending on a particular context for a specific movement (e.g., feeding or competition) and the amount of energy required[18]. Designing multi-latched LaMSA systems in engineered materials is appealing yet largely unexplored due to the complexity of integrating multiple components into a single material platform.

Here, we report a bistable magneto-elastic shell device integrated with a dual-latch mechanism, an intrinsic mechanical latch and an external magnetic latch. Differential deswelling through the thickness direction acts as a "motor", converting physicochemical energy into stored elastic energy in the gel, which is subsequently released through a snap-through instability, generating a power-dense actuation. The selective activation of a second, external magnetic latch enables distinct actuation modes, allowing for exceptional control of snapping onset and a significant (greater than 20-fold) power amplification of triggered motion, thereby enhancing adaptability to various application scenarios.

The magneto-elastic gel device consists of a flexible polymer matrix embedded with radially aligned hard NbFeB particles and is swollen in acetone (**Fig. 1a**). The shell geometry of the gel device, characterized by the curvature ($\kappa = 1/r$), central angle ($\theta$), and thickness ($h$), determines the bistable nature and the intrinsic mechanical latch capacity of the gel. Additionally, the shell geometry contributes to the formation of an untethered, autonomous "motor" to store energy in the elastic gel (**Fig. 1b**). Specifically, as the swollen gel undergoes free evaporation in ambient air, the faster evaporation rate at the convex side than at the concave side drives a strain gradient along the thickness direction. This resultant through-thickness gradient of internal stresses favors inversion of the spherical curvature, and its buildup stores elastic energy in the gel. Once the internal stresses reach the critical threshold ($\sigma = \sigma_c$) set by geometric and material properties, the snap-through instability is triggered immediately, releasing the stored energy (**Fig. 1c**). Prior to this, the through thickness stress gradients are insufficient to induce snapping, which we define as the subcritical stage ($\sigma < \sigma_c$). In contrast, when exposed to a magnetic field, the external magnetic latch is activated. The unique radial NbFeB alignment leads to a magnetic torque that favors the initial convex shell curvature and resists snapping (**Fig. 1c**). This allows the elastic energy to further accumulate even beyond the intrinsic critical threshold of the free shell. Once the magnetic field is removed, the accumulated energy is rapidly released, driving a power-amplified snap-through instability. We define this enhanced snapping as *supercritical snapping* ($\sigma > \sigma_c$), in analogy with the supercritical behavior in thermodynamics, in which a system evolves continuously between states, unimpeded by the existence of a barrier. With the advent of a magnetic latch, the snapping onset and energy output can be far more precisely controlled based on the timing of the magnetic field removal and the duration of its application, respectively. In both critical and supercritical snapping events, the released energy is used to launch the device via the reaction force with the substrate during snapping (**Fig. 1d**).

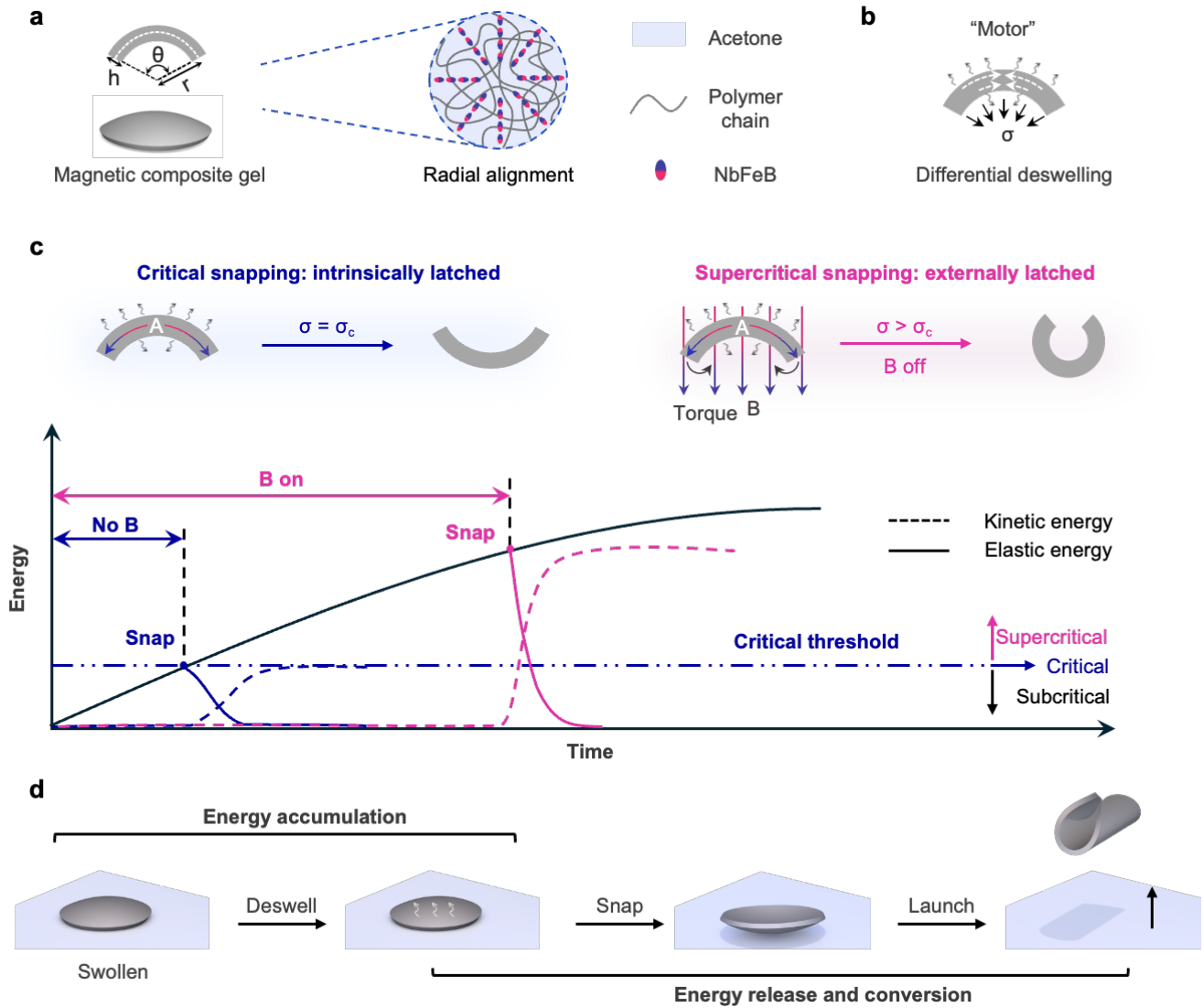

**Fig. 1 | Schematic illustration of the dual-latched magneto-elastic gel device. a**, Dome shaped magnetic composite gel with radial alignment. **b**, Differential deswelling acting as a motor, leading to internal stresses. **c**, Mechanism of internally latched critical snapping (geometric latch) and externally latched supercritical snapping (magnetic latch) as well as corresponding energy evolution. **d**, Launching of the gel by the reaction force from the substrate after snapping.

**Fabrication of the magneto-elastic gel device**

The polyethylene glycol (PEG) polymer matrix is crosslinked via a photocurable thiol-ene click chemistry, and we selected the formula with 1.2 mol% crosslinker to ensure a flexible network (**Supplementary Fig. 1a, b**). A radially magnetized ring magnet is used to radially align the magnetic NbFeB particles (~5 μm, **Supplementary Fig. 1c**) before they are immobilized during crosslinking of the polymer network (**Supplementary Fig. 2a**). However, the external magnetic field is not uniform, with greater strength near the ring magnet, causing the particles to accumulate at the edge (**Supplementary Fig. 2b, c**). These accumulated particles act as photo absorbers, in turn reducing the monomer conversion rate more at the outer radii than at the center of the disc. Removal of unreacted monomers then leads to relative shrinkage at the outer perimeter of the disc corresponding to a non-Euclidean metric that prefers a positively curved spherical shape[19]. In this initial state, the flat disc is equally unstable to convex or concave buckling. The gel device is obtained by further swelling the dome in a volatile solvent, such as acetone used here, until the swelling equilibrium is reached.

Magnetic particle loading plays an important role in making the magneto-elastic gel device. To investigate its influence, a series of composites with varying particle loadings were prepared to explore the effects on monomer conversion, mechanical properties, and geometrical parameters of the obtained gel device.

To quantify the monomer conversion rate at various particle loadings, we measured the mass loss ratio ($\delta_c$) of the as cured composite after rinsing in toluene to remove unreacted monomers (**Supplementary Fig. 3a**). However, the particles might also be removed as the polymer network swells and expands. By analyzing the mass loading of the composite before ($L_0$) and after ($L'$) monomer removal with thermogravimetric analysis (TGA), we confirmed that only unreacted monomers were washed off and monomer conversion rates were then calculated by $\frac{1-L'}{1-L_0}(1-\delta_c)$ (see **Supplementary Section 1, Supplementary Fig. 3b**). The monomer conversion result shows that the photochemistry is inhibited when the particle loading is high enough to absorb a substantial amount of light (**Supplementary Fig. 3c**). The incomplete reaction is also verified by the decreased modulus, increased hysteresis, and increased loss modulus of the composite with particle loading (**Supplementary Fig. 4**). The curvature and the central angle of the 3D shells both increase with particle loading, indicating a more curved and deeper shell at high particle loading (**Supplementary Fig. 5a**). However, excessively high particle loading severely hinders the polymerization reaction, preventing the polymer network from effectively trapping the particles. Thus, we selected 36 wt% particle loading and fabricated shells from flat discs of height ($h_{dry}$ = 0.35 mm) and with different radii ($R_{dry}$ = 4 mm, 5 mm, 6 mm, **Supplementary Fig. 5b**). Acetone was selected due to its high swelling ratio at equilibrium ( $\varepsilon_{eq} = \frac{R_{eq} - R_{dry}}{R_{dry}} \times 100\% \approx 0.4$ ) and rapid evaporation speed (**Supplementary Fig. 6**). The diffusivity ($D$) in acetone was measured to be 0.14 x 10$^{-9}$ m$^2$/s by indentation[20] (**see Supplementary Section 2**). Notably, at 36 wt% particle loading, the particle alignment is sufficiently robust to withstand repeated swelling and deswelling cycles, an essential prerequisite for subsequent snapping and launching experiments.

**Dual latched snapping and jumping**

Here, we selected a gel device with 36 wt% particle loading, fabricated from a flat disc with a 5 mm radius ($R_{eq} = R_{dry}(1 + \varepsilon_{eq}) = 7\ mm$, $h_{eq} = h_{dry}(1 + \varepsilon_{eq}) = 0.5\ mm$) for the following snapping and launching experiments. During free deswelling, the evolving internal stress gradients reduce the curvature but do not invert the shell due to the intrinsic latch provided by the finite elastic energy barrier between initial and inverted shapes. At approximately 5 seconds (**Fig. 2a)**, the internal stresses surpass the critical threshold, triggering a snapping event that propel the gel off the substrate and into the air. When subjected to an external magnetic field ($B$ = 130 mT) generated by an electromagnet, a secondary, embedded magnetic latch is activated, which is strong enough to preserve the initial curvature and suppress snapping throughout the entire deswelling process until the gel is fully dry (**Fig. 2a**). Strategically removing the magnetic field results in a more powerful, supercritical snapping and a higher jump height. An optimal holding time of 90 seconds under the applied magnetic field maximizes the jump height to 117 mm, compared to only 11 mm without the magnetic field (**Fig. 2b**). Snapshots in **Fig. 2c** illustrate the jumping performance under both snapping modes (see **Supplementary Video 1**). The takeoff velocity reaches 1.7 m/s during supercritical snapping with the assistance of the magnetic field, compared to 0.5 m/s without it (**Fig. 2d**). The energy output was calculated from the jump height, yielding a peak value of 68 µJ and a peak energy density of 1163 µJ/g (**Supplementary Fig. 7**).

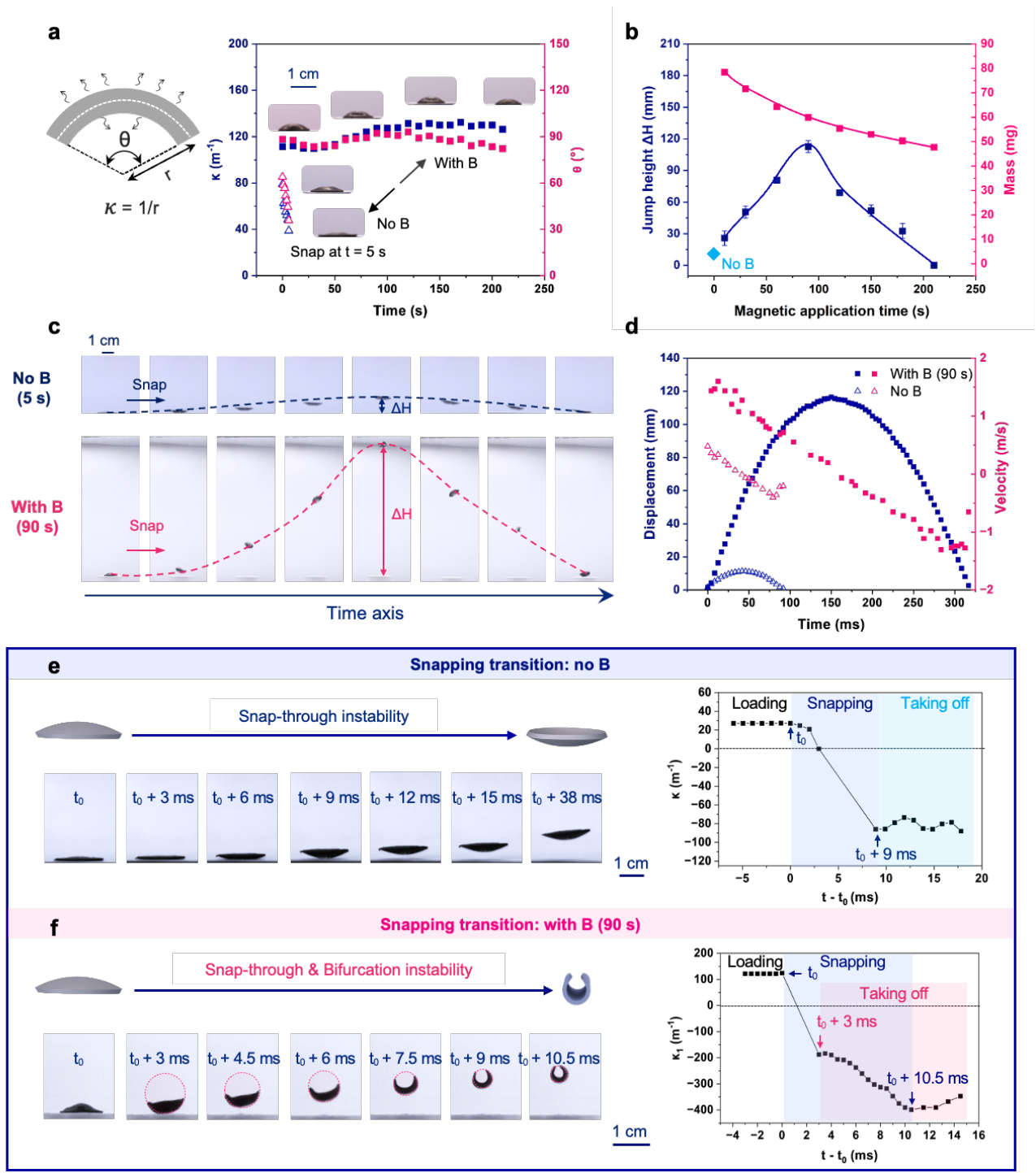

**Fig. 2 | Dual latched snapping and jumping. a**, Evolution of geometric parameters of the gel over time during deswelling. **b**, Jump height and mass of the gel as a function of magnetic field application time. **c**, Snapshots of the jump without magnetic field (top) and when the magnetic field is applied for 90 s (bottom). **d**, Corresponding displacement/velocity-time plots in (**c**). **e**, **f**, High-speed snapshots (captured at 20,000 fps) and corresponding curvature evolution during snapping transition: without magnetic field (**e**) and when the magnetic field is applied for 90 s (**f**).

To further quantify the performance, the snap-through transition was recorded using a high-speed camera (see **Supplementary Video 2 and 3**) with a frame rate of 20,000 frames per second (fps). During critical snapping, the transition and takeoff occur within 9 ms ($t_{\text{transition}}$), yielding a power density of 10 W/kg, calculated using $P_m = g\Delta H / t_{\text{transition}}$ (**Fig.2e**). In contrast, for supercritical snapping, shape inversion is completed in just 3 ms, increasing the power density to 387 W/kg (**Fig. 2f**). Notably, the snapping continues even after takeoff during supercritical snapping, eventually forming a cylindrical shell, in contrast to its inverted spherical configuration observed in critical snapping. This symmetry breaking indicates that the gel undergoes not only a snap-through instability but also a shape bifurcation[21-23].

**Supercritical snapping and shape bifurcation**

Here, we describe a simplified theoretical model that illuminates how the mechanisms underlying supercritical snapping also drive bifurcation of the post-snap shape. The driving force for spontaneous snap-through of the gel is the development of the strain gradient along the thickness direction, induced by solvent evaporation from the curved gel. Initially, the fully swollen gel exhibits a uniform solvent distribution. As solvent evaporates, the larger surface area at the convex top of the shell leads to more net deswelling on that side relative to the concave bottom, giving rise to a relative deficit of solvent near the top and a peak of the concentration closer to the bottom (**Fig. 3a**, **Supplementary Fig. 8**). This solvent concentration imbalance favors relative contraction of the upper surface compared to the bottom, generating a bending moment directed toward shell inversion. At sufficiently long times, all the solvent evaporates from the shell, and the through-thickness gradient disappears, implying that this "inward" bending moment also vanishes. Assuming that local strains are proportional to through-thickness solvent excess and that the solvent concentration follows Fick's law of diffusion from a spherical shell (see **Supplementary Section 2**), it is straightforward to predict the dynamics of the inward bending moment as a function of time, showing that it indeed gives rise to a peak at time $0.055(h^2/D)$ – where $D$ is the solvent diffusion constant in the gel – and decays exponentially beyond that time (**Supplementary Fig. 8**).

To model the shape mechanics of this scenario, we adopt the elastic energy of shallow-shell model, with an elastic energy that can be written in terms of its two approximately constant principal curvatures ($\kappa_x, \kappa_y$), or equivalently its Gaussian ($G$) and mean ($H$) curvatures,

$$U_{\text{elastic}} = U_{\text{stretch}} + U_{\text{bend}} = \frac{\pi R^6 Y}{384}(G-G_0)^2 + \frac{\pi R^2 B}{2}(2H-2H_0)^2 \tag{1}$$

where here $Y = Eh$ and $B = \frac{Eh^3}{12(1-\nu^2)}$ are in-plane stretching and bending moduli for gel with Young's modulus $E$ and Poisson's ratio $\nu$. The preferred Gaussian curvature $G_0 > 0$ arises primarily due to outer perimeter contraction following polymerization and we assume it to be roughly constant during the deswelling process, an approximation we later verify by comparison with observations. On the other hand, the preferred mean curvature $H_0$ is generated by through-thickness gradients of solvent-induced swelling,

$$H_0(t) \propto \frac{1}{Eh^3} \int_{-\frac{h}{2}}^{\frac{h}{2}} \Pi(z) z \, dz \tag{2}$$

where $\Pi(z)$ is the solvent pressure at distance $z$ from the mid-layer of the gel, which is proportional to local solvent concentration according to standard (e.g. Flory-Rehner) descriptions of gels. Due to the swelling dynamics described above, $H_0(t=0) = 0$ in the initial, uniformly swollen state and increases in magnitude with time as the through-thickness solvent gradient develops, with a sign *opposite* to its current mean curvature, before reaching a maximum at $t = 0.055(h^2/D)$. The preferred mean curvature, $H_0(t)$, then decays back to zero at long times, akin to similar transient bending observed for asymmetric swelling of initially planar films. We can consider the shape thermodynamics under this situation of increasing bending drive to invert the shell via a simplified reduced elastic energy,

$$\frac{U_{\text{elastic}}}{U_0} = \frac{1}{2}(\bar{G}-\bar{G}_0)^2 + \frac{\bar{h}^2}{2}(2\bar{H}-2\bar{H}_0)^2 \tag{3}$$

where $\bar{H} = HR$ and $\bar{G} = GR^2$ dimensionless curvatures, $\bar{h} = \frac{4}{\sqrt{1-\nu^2}}\left(\frac{h}{R}\right)$ is dimensionless thickness and $U_0 = EV/192$ is the characteristic elastic energy scale for the gel.

The reduced elastic energy in equation (3) defines 2D energy landscape for the shell shape in terms of principal curvatures, $\bar{\kappa}_x$ and $\bar{\kappa}_y$, defined in terms of preferred curvatures, $\bar{H}_0$ and $\bar{G}_0$, and $\bar{h}$ which characterizes relative ratio of bend to stretch stiffness in the gel. We model the snapping behavior by considering the evolution of the landscape – its local minima and saddle-points – as function of increasing mean curvature, assuming fixed Gaussian curvature and thickness, considering a shell that begins in a convex and spherical local minima, $\bar{H} > 0$ with $\bar{\kappa}_x = \bar{\kappa}_y$. As described in **Supplementary Section** 3, the shell landscape is described by two types of transition, depending on two preferred mean-curvature thresholds. The first $|\bar{H}_0|_c = \frac{(\bar{G}_0 - 2\bar{h}^2)^{3/2}}{\sqrt{27\bar{h}^2}}$ describes the threshold curvature below which a finite energy barrier separates a metastable local convex ($\bar{H} > 0$) minimum from the inverted ($\bar{H} < 0$) global energy minimum state(s). In deswelling free shells, once the $|\bar{H}_0|$ grows to reach $|\bar{H}_0|_c$ it undergoes *critical snapping*, rapidly buckling from its saddle point to an inverted shape. A second geometric threshold $|\bar{H}_0|_b = \sqrt{\bar{G}_0}$ determines the symmetry-breaking behavior of the energy landscape. When $|\bar{H}_0| > |\bar{H}_0|_b$ the global energy minima are described by non-axisymmetric two shapes $\bar{\kappa}_x = \left(\bar{H}_0 \pm \sqrt{\bar{H}_0^2 - \bar{G}_0}\right), \bar{\kappa}_y = \left(\bar{H}_0 \mp \sqrt{\bar{H}_0^2 - \bar{G}_0}\right)$, which satisfy target shape of both *stretching* and *bending* energy perfectly (i.e. $U_{\text{elastic}} = 0$). When $|\bar{H}_0| < |\bar{H}_0|_b$ the global minimum is *geometrically frustrated* and retains an axisymmetric, inverted shape with residual elastic energy (i.e. $U_{\text{elastic}} > 0$), if still somewhat lower than the metastable convex state **(Supplementary Fig. 9)**.

Based on this survey of the shell energy landscape, it is straightforward to understand the observed snapping behavior and performance deswelling gels. First, our gel devices where scaled curvature and thickness are in the range of $\bar{G}_0 \approx 0.35 - 0.65$ and $\bar{h} \approx 0.33$ correspond to mean curvature thresholds

$|\bar{H}_0|_c \approx 0.10 - 0.50 < |\bar{H}_0|_b \approx 0.60 - 0.80$. Hence, in the absence of an external field, deswelling proceeds until $|\bar{H}_0|$ first reaches the *critical snapping* threshold. Since this is well before shape bending moment to induce shape bifurcation, these shells invert snap to frustrated shape, that retains axisymmetry and residual elastic energy. This residual elastic energy in the snap-through shape accounts for the considerably lower kinetic energy of the critical snap events.

In the presence of the downward magnetic field, i.e. the second external latch, the alignment of the magnetic particles couples the mean curvature of the shell, leading to a total potential energy (**see Supplementary section 3**)

$$\frac{U_{\text{total}}}{U_0} = \frac{U_{\text{elastic}}}{U_0} - \frac{64\rho_m B}{Eh}(2\bar{H}) \qquad (4)$$

where $\rho_m$ is the magnetic moment density per surface area of the embedded particles ($J/Tm^2$), $B$ is the external magnetic field strength ($T$). This applied field stabilizes convex shapes, even as the deswelling induced through-thickness gradient drives the gel towards inversion. This can be characterized as an effective shift of the preferred mean curvature towards positive, convex shapes by $\Delta\bar{H}_0(B) = \frac{2\rho_m B R^2(1-\nu^2)}{Eh^3} > 0$. Hence the applied field reduces the overall magnitude of $|\bar{H}_0(t) + \Delta\bar{H}_0(B)|$, keeping the gel below the critical threshold of $|\bar{H}_0|_c$ even as the solvent-induced bending moment grows to exceed that threshold (i.e. $\bar{H}_0(t) < -|\bar{H}_0|_c$). Then, as the field is removed ($B = 0$) the gel is immediately quenched into a *supercritical state* where its current convex shape is deep in the unstable regime, leading to quick inversion to stable concave shapes (**Supplementary Fig. 10**). We refer to this scenario as *supercritical snapping*, since the gel is released from a point far beyond its own intrinsic stability limits. Holding with the applied field for sufficiently long times ultimately allows the swelling gradient to climb above the second bifurcation threshold $|\bar{H}_0(t)| \geq |\bar{H}_0|_b$ allowing the snap-through to release all residual energy as it assumes an axisymmetric equilibrium shape.

We test this scenario, by noting that when the final, post-snap gels adopt shapes for which $H^2/G > 1$ the shells are unfrustrated, and we can infer the preferred mean and Gaussian curvatures from their final shapes. Experimental data shows that with increasing magnetic field application time, the gel shape after snapping transitions from a spherical cap to a cylindrical shape, attaining peak curvature, and then gradually relaxes back to a spherical form (**Fig. 3b, Supplementary Fig. 11**). This trend is reflected by $H_0^2/G_0$, which grows greater than 1, and eventually returns to 1 over longer hold times (**Fig. 3c**). Larger shells penetrate deeper into the bifurcation zone. The peak value, representing the maximum strain gradient, is observed at magnetic holding time of 90 s, which aligns well with the predicted time scale $0.055(h^2/D)$ (**Supplementary Fig. 11c**). Notably, 90 s also corresponds to the time at which the maximum jumping height is achieved, indicating a strong correlation between the strain gradient and energy output.

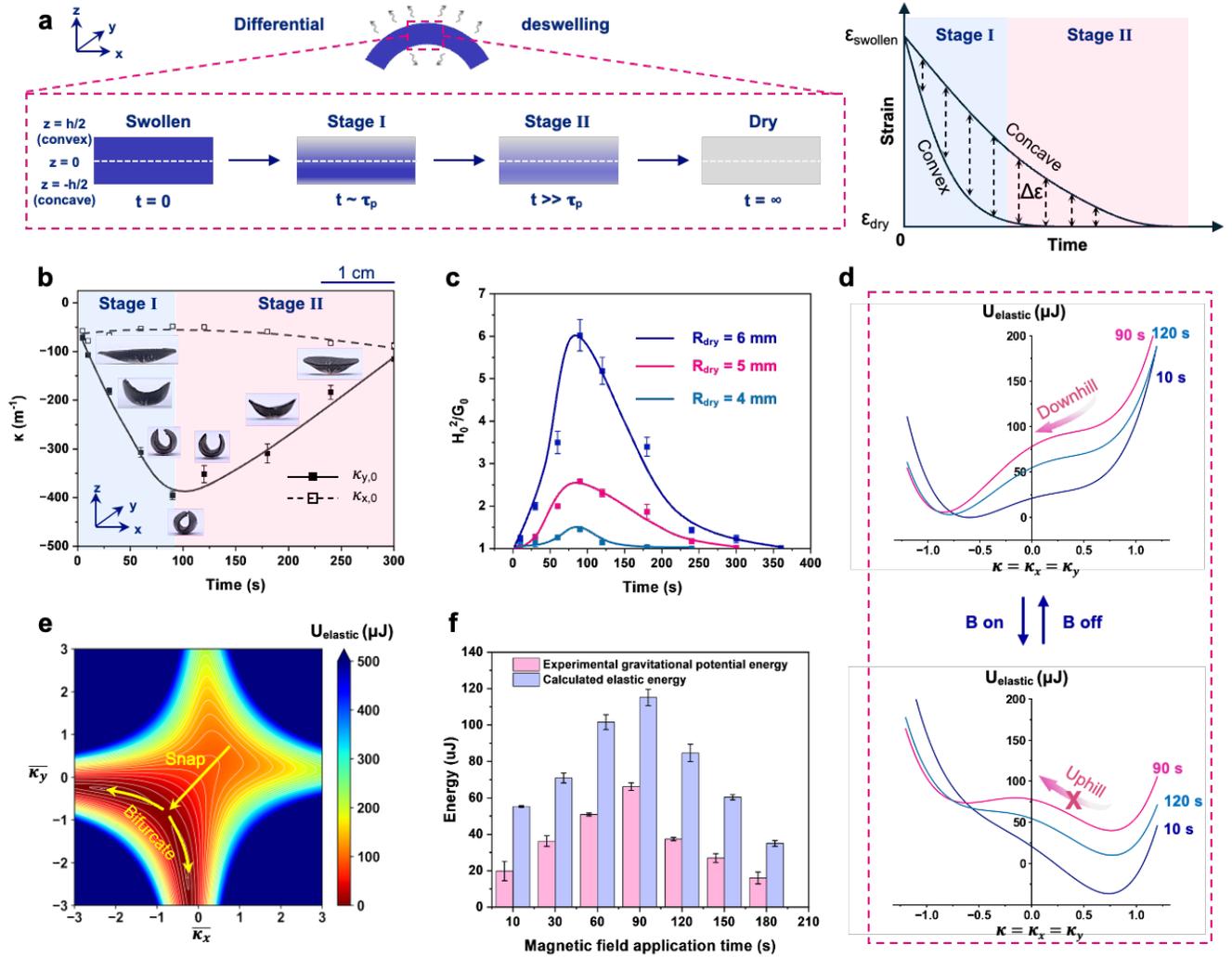

**Fig. 3 | Theory of supercritical snapping and bifurcation a**, Schematic illustration of strain gradient evolution during deswelling. $\tau_p$ is the poroelastic time, $\varepsilon_{dry}(0)$ is the strain at dry state and $\varepsilon_{eq}$ is the strain at equilibrium swollen state. **b**, Evolution of principal curvatures of post-snapping gels for different magnetic application durations. **c**, Evolution of $H_0^2/G_0$ of gels with different sizes. **d**, The stabilization effect of supercritical state with the magnetic field. **e**, Top-down plot of energy landscape for a bifurcated gel (magnetic field is applied for 90 s. **f**, Comparison of calculated elastic energy and experimental energy output.

To illustrate the energetic function of the dual latch system, we input all experimentally measured geometric ($R$, $h$, $\bar{H}_0$ and $\bar{G}_0$) and mechanical parameters ($E$, $\nu$) into equation (3) and plot the energy landscapes for various durations of magnetic field application (**Supplementary Fig. 12**). We find supercritical snapping that surpasses the intrinsic energy barrier in the absence of a magnetic field and are successfully stabilized when the magnetic field is applied (**Fig. 3d**). The stronger the magnetic field, the more pronounced the energy landscape shift (**Supplementary Fig. 13**). After removing the magnetic field, supercritical snapping occurs followed by splitting into two energy minima as expected (**Fig. 3e**). As the strain gradient gradually decreases over time, the gel relaxes, and the energy minima converge back to a single conjugate point (**Supplementary Fig. 12**). The elastic energy stored in the gel under varying magnetic field application times was calculated using equation. (3) and compared with the experimentally measured gravitational energy output (**Fig. 3f**). While we find strong, fit-free agreement for energy release from this simplified model over broad parameter ranges studied, some quantitative discrepancies remain, which are likely attributed to additional dissipation mechanisms in the conversion from stored elastic to kinetic energy: from elastic energy to kinetic energy, and subsequently to gravitational potential energy.

**Two pathways of energy storage and release**

Based on the discussions above, critical snapping releases less energy than supercritical snapping. However, the preserved structural symmetry following critical snapping allows the gel to autonomously undergo consecutive jumping events—provided it flips upon landing—continuing this behavior until it is fully dried (**Supplementary Fig. 14**). Interestingly, when the total energy output from all consecutive jumps is summed, it becomes comparable to that of the supercritical snapping (**Supplementary Fig. 15**). The fundamental distinction lies in the mechanisms of energy storage and release: critical snapping dissipates energy over multiple smaller cycles, whereas supercritical snapping accumulates energy and releases it in a single, forceful event. These contrasting energy release modes are illustrated in **Fig. 4a** and **4b**. When placed on a slope (6.5°), a lateral launching of the gel is achieved. In critical snapping mode, the gel reaches the target location with multiple small steps, while in supercritical snapping mode, it achieves the same destination with a single large jump (see **Supplementary Video 4**). Overall, the selective activation of the external magnetic latch enables on-demand energy storage and release, mimicking the adaptive responses of living organisms to varying environmental conditions (**Supplementary Fig. 16**).

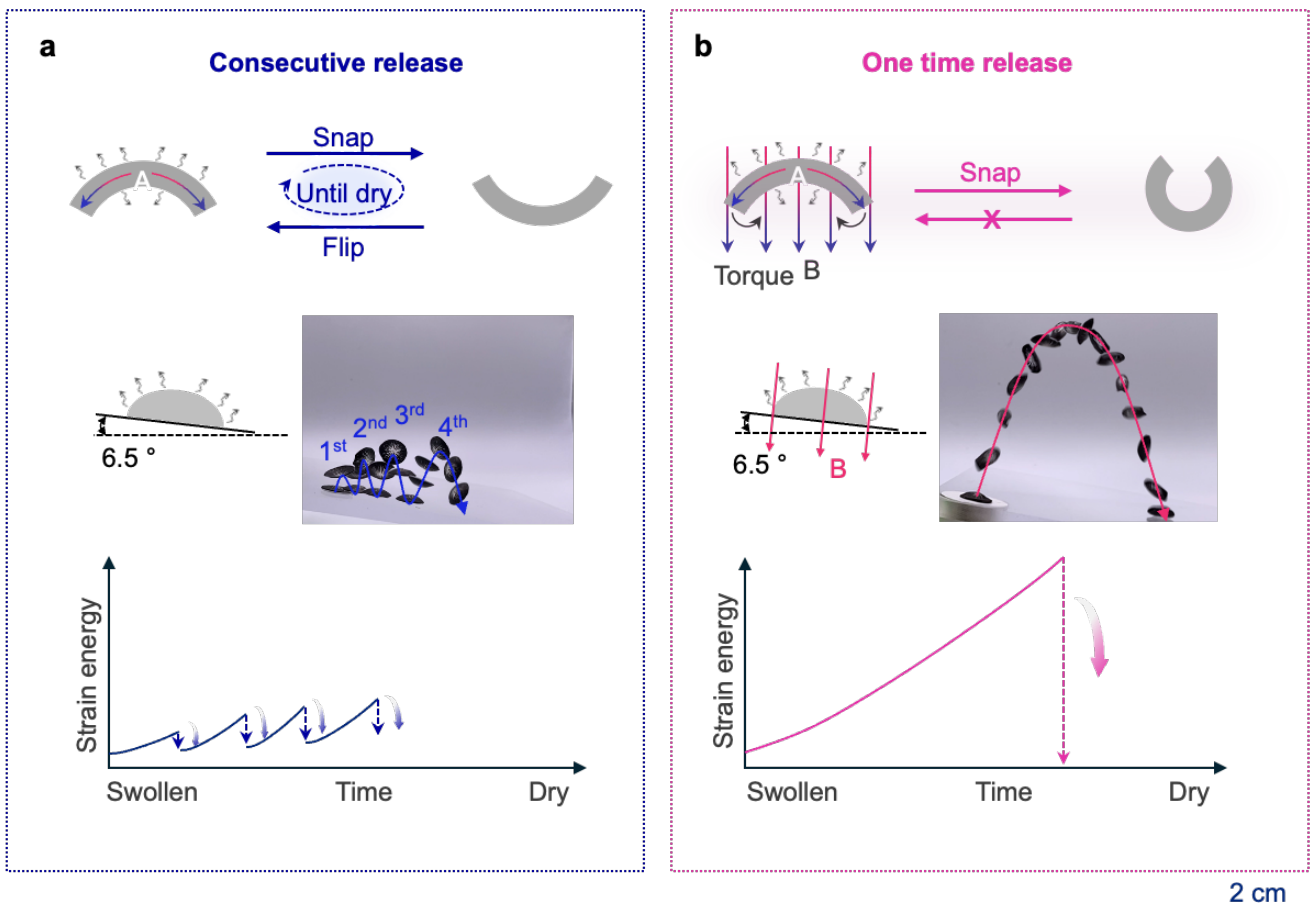

**Fig. 4 | Control of energy storage and release pathways. a**, The energy storage and release in multiple small cycles in critical snapping pathway. **b**, The energy storage and release in a single powerful burst in supercritical snapping pathway.

## Control of launch direction

In addition to enhancing energy output, activation of the magnetic latch offers improved control over the launch direction of the gel, enabling trajectories that are unattainable without the magnetic latch. When exposed to an external magnetic field, the gel experiences both magnetic torque, which resists snapping to accumulate more energy, and a magnetic attractive force generated by the uneven magnetic field. This attractive force secures the gel to the substrate without interfering with the deswelling gradient. As a result, the substrate can point in any direction, allowing the gel to launch in the opposite direction when the magnetic field is turned off (see **Supplementary Video 5)**. **Fig. 5a** demonstrates launches ranging from upward to even downward when a magnetic field is applied for 60 seconds. The downward launch is distinct from a simple gravity-driven fall, as the gel has an initial takeoff velocity of 1.0 m/s (**Fig. 5b**). **Fig. 5c** shows velocity components ($V_x$, $V_y$), where the consistent resultant velocity confirms no energy loss when launch direction varies. If the direction is fixed, the launch can be further controlled by adjusting the magnetic application time (**Fig. 5d, Supplementary Fig. 17**). This directional launch capability provides the gel with substantial versatility for soft robotics applications. It can enable a basketball shot from different locations on a toy-scale court (**Fig. 5e, Supplementary Video 6**), or execute in-plane launches, such as in a game of billiards (**Fig. 5f, Supplementary Video 7**). Collisions transfer the energy to the beads (**Fig. 5g, Supplementary Video 8**). The device can also overcome obstacles by either slipping through small openings at the bottom or jumping over them from above (**Fig. 5h, Supplementary Video 9**).

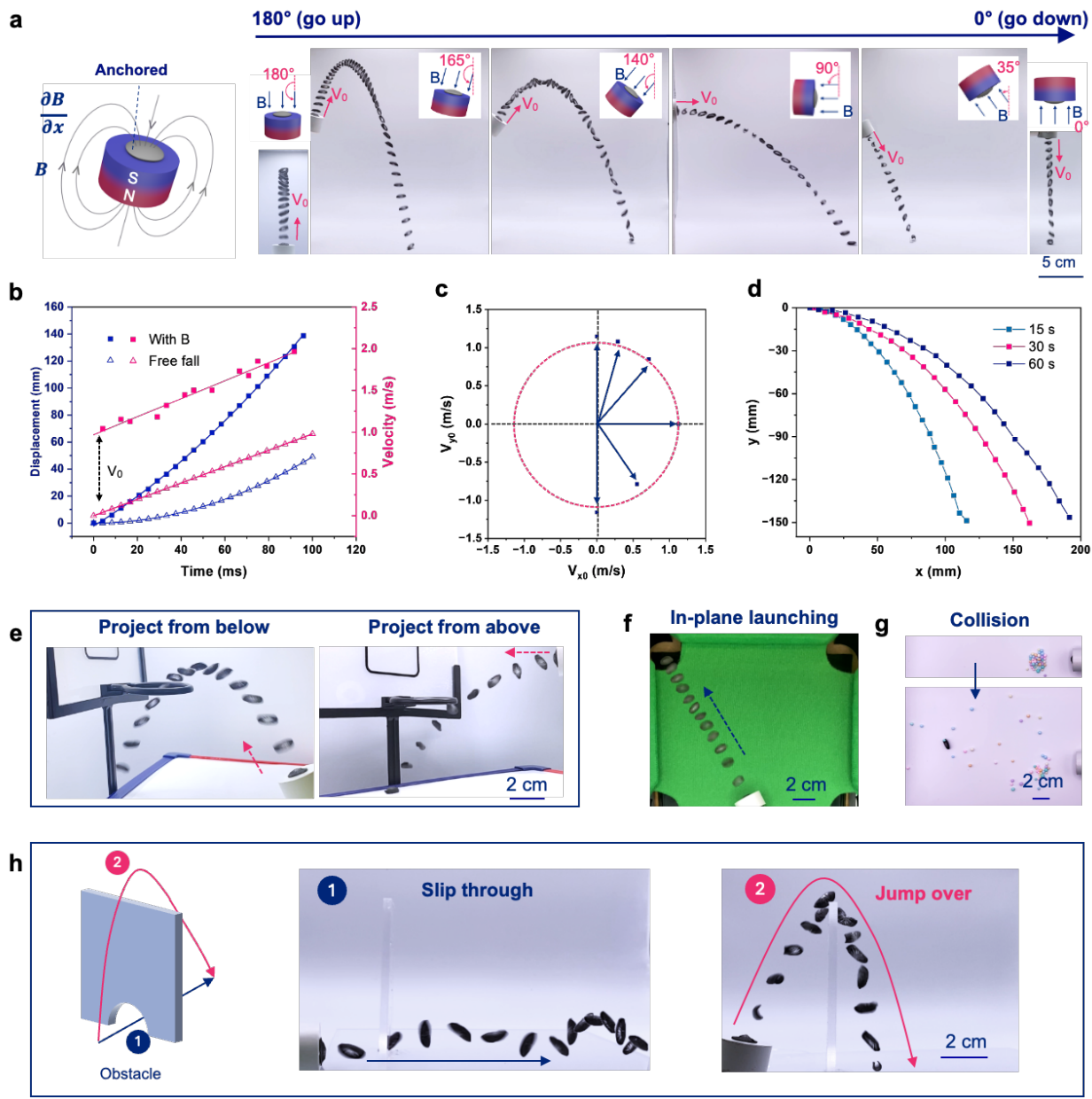

**Fig. 5 | Control of launch direction. a**, Anchoring mechanism and directional launching when magnetic field is applied for 60 s. **b**, Displacement/velocity-time plots of downward launching and gravity-driven free fall. **c**, Takeoff velocity and its x, y components when launching to different directions. **d**, Time controlled launching with a fixed direction of 90°. **e**, Launching the gel to into the basket from different angles. **f**, In plane launching the gel into a hole. **g**, Collison with the beads to transfer energy. **h**, Slipping through or jumping over an obstacle.

## Conclusion

We have presented the design principles of a magneto-elastic gel device embedded with dual latches that can be selectively activated to mediate the snapping pathways and launching of the gel. Our approach takes lessons from integrated LaMSA systems in nature to adapt to different application scenarios. We demonstrate the ability to generate supercritical snap-through actions, in which actuation is delayed well beyond the critical threshold, significantly enhancing power output. Additionally, the simple yet effective anchor mechanism enables the gels to launch in arbitrary directions, which provides the gels with substantial versatility for soft robotics applications.

Our theoretical framework describes the underlying physics of dual-latch materials systems and broadly applies to general LaMSA systems, opening pathways for quantitatively designing multiple latch systems to achieve desired outputs. We envision that the quantitative mapping of performance desires or needs can be overlaid with this theoretical mapping to aid the design of LaMSA materials and associated devices. Furthermore, our theory describes how additional latches, when coupled with mechanical instabilities, allow supercritical instabilities to be achieved. To the best of our knowledge, this is the first description and demonstration of such states, which should have broad appeal across fields ranging from biology to mechanics to materials. It is reasonable to expect this dual latch framework can apply to (de)swelling induced instabilities applied to additional geometries (e.g. compressive or torsional rod buckling), raising new questions about role of this geometric mechanics in controlling performance (e.g. power output). We further envision that these design concepts can be extended to other material systems, for example, using a phase-change material as an internal phase-locking latch paired with a geometric latch to achieve intrinsically autonomous supercritical snapping. Collectively, such advances are predicted to impact the development of robotic devices, adaptable prosthetics, and advanced manufacturing methods.

homogenous swelling. *Soft Matter*, **7**, 5188-5193 (2011).

## Materials

Diethylene glycol divinyl ether (DEGDE), 3,6-dioxa-1,8-octanedithiol (DODT), Pentaerythritol tetrakis (3-mercaptopropionate) (PETMP, crosslinker) and 2,2-dimethoxy-2-phenylacetophenone (DMPA, photo-initiator) were sourced from Sigma-Aldrich. All the chemicals and solvents were used as received without further purification. Magnetic (NbFeB) particles were purchased from Neo Magnequench Inc. (20441-089, 5μm).

## Preparation of the polymer films without magnetic particles

The sample with 1.2% (molar ratio) crosslinker was selected to illustrate the synthesis protocol. In detail, 2 g DODT, 2.33 g DEGDE, and 0.02 g DMPA (0.5 wt%) were sequentially added to a vial and stirred until a homogeneous mixture was achieved. The mixture was then exposed to a house-built UV illumination system (intensity: 10 mW/cm$^2$) for 3 minutes, resulting in the formation of extended linear polymer chains terminated with vinyl groups. Next, 0.13 g PETMP and an additional 0.02 g DMPA (0.5 wt%) were added to the mixture. After stirring and degassing, the mixture was poured into a glass cell, sandwiched between 0.3 mm PDMS spacers, and exposed to UV light for 4 minutes, 2 minutes on each side (top and bottom). The transparent polymer films were obtained after allowing the toluene to evaporate in a fume hood for 2 hours. Films with different crosslinking densities were prepared using the same procedure, with variations made only to the formulations.

## Preparation of randomly dispersed magnetic composite

The sample with 1.2% (molar ratio) crosslinker and 36 wt% particle loading was selected to illustrate the synthesis protocol. First, 2 g DODT, 2.33 g DEGDE, and 0.02 g DMPA (0.5 wt%) were added to a glass vial and exposed to UV light for 3 minutes. The initial reaction between DODT and DEGDE increases the viscosity of the mixture, which is crucial for ensuring proper dispersion of the magnetic particles and preventing aggregation. Next, 0.13 g PETMP, 0.02 g DMPA (0.5 wt%), 1 g toluene (to slightly reduce viscosity for better mixing), and NbFeB particles were added. Note that the NbFeB particles were magnetized using impulse magnetic fields (2.7 T) generated by an impulse magnetizer (IM-10-30, ASC Scientific), which imparted magnetic polarities to the particles. The slurry was manually stirred for 5 minutes, followed by degassing under vacuum. It was then poured into a glass cell, sandwiched with a 0.3 mm PDMS spacer, and exposed to UV light for 4 minutes, 2 minutes on each side (top and bottom). After removing the toluene by leaving the sample in a fume hood for 2 hours, the randomly dispersed magnetic composites were obtained. Samples with different particle loadings were prepared following the same procedure, with variations only in the formulations.

**Preparation of magnetic composite shells with radial alignments**

The procedure is similar to that used for randomly dispersed magnetic composites. After preparing the magnetic slurry, it was poured into a PMMA cell (disc-shaped and cut using a laser cutter) sandwiched between two PDMS spacers (0.3 mm thick, also disc-shaped and cut by a laser cutter: Universal Laser Systems VLS3.50). The sample was then exposed to a radial magnetic field generated by a radially magnetized ring magnet, causing the magnetic particles to align along the external magnetic field. This alignment was fixed through crosslinking by exposing the sample to UV light for 4 minutes, 2 minutes on each side (top and bottom). To remove unreacted monomers, the sample was washed three times with toluene, each wash lasting 2 hours. After drying, magnetic composite shells with radial alignment were successfully obtained.

**Snapping and jumping of composite gel without magnetic field**

After swelling to equilibrium in acetone, the composite gel was taken out and any excess acetone on its surface was wiped off. The gel was then placed on a smooth substrate with the convex side facing upward. As internal stresses developed, it suddenly snapped and jumped at some time. When it landed back on the substrate, again with the convex side facing upward, it was primed to snap and jump autonomously for the second time without additional swelling in acetone. Otherwise, manual flipping was required to ensure the convex side was always facing up to ensure a snap-induced jump. The snapping and jumping repeated several times until the internal swelling gradient of the gel was no longer great enough to generate a critical stress.

**Snapping and jumping of composite gel with the assistance of magnetic field**

After swelling to equilibrium in acetone, the composite gel was removed and excess acetone on its surface was wiped off. The gel was then placed on a commercial cylindrical electromagnet (Uxcell MF-P25/20, purchased from Amazon) which was connected to a DC power supply to generate a downward external magnetic field. After deswelling for a certain time, the magnetic field was turned off, causing the gel to snap and jump to various heights.

**Measurement of diffusivity in a polyethylene glycol (PEG) polymer matrix**

The diffusivity of PEG polymer matrix was measured following the procedures in previous literature[20]. In summary, a solvent bath was constructed and filled with acetone. A cylindrical puck of PEG (R ~ 12 mm, h ~ 7.5 mm) was submerged within the solvent bath. A Texture Analyzer (TA-XT PLUS, Stable Microsystems) fitted with a spherical probe (R = 12.7 mm) was indented 0.3 mm into the cylindrical sample at 0.05 mm/s and the resultant force was measured over time as solvent transports from the compressed region (See **Supplementary Section 2** and **Fig. S6d** for the analysis and results of the diffusivity measurement).

**Material characterizations**

The particle size was confirmed by Scanning Electron Microscopy (SEM, FEI Magellan 400 XHR-SEM) The radial particle alignment was confirmed by Optical Microscopy. The mechanical tests were performed using a Texture Analyzer (TA-XT PLUS, Stable Micro Systems) at 25 ºC with a speed of 1 mm/s and

Dynamic Mechanical Analyzer (TA Discovery DMA 850) in a tension mode. The particle loadings before and after washing were measured by Thermogravimetric Analysis (TGA Q500). The photos of snapping transitions were taken by high-speed camera (20000 fps), others were taken from iPhone 11 in the slow-motion mode (240 fps).

## Author contributions

X.M.X and A.J.C. conceived the idea. X.M.X conducted the experiments. N. A. M performed the diffusivity measurements. G.M.G. contributed to the modeling and theory support. X.M.X, G.M.G and A.J.C. wrote the manuscript. All co-authors performed data analysis and editing of the manuscript.

**Competing interest:** The authors declare no competing interests.

## Acknowledgements

This research was funded by the University of Massachusetts Amherst and the U. S. Army Research Laboratory under contract/grant number W911NF-23-2-0022. G.M.G. acknowledges support from the US National Science Foundation through Award DMR 2349818. We are grateful for the insightful discussions with members of the Crosby research group. We thank Xuanhe Zhao and Jaehun Choe for their assistance with particle magnetization, and Xuchen Gan for his support with Python coding.